\documentclass[conference]{IEEEtran}
\IEEEoverridecommandlockouts

\usepackage{cite}
\usepackage{url}
\usepackage{amsmath,amssymb,amsfonts}
\usepackage{algorithmic}
\usepackage{graphicx}
\usepackage{textcomp}
\usepackage{multicol}
\usepackage{multirow}
\usepackage{xcolor}
\def\BibTeX{{\rm B\kern-.05em{\sc i\kern-.025em b}\kern-.08em
    T\kern-.1667em\lower.7ex\hbox{E}\kern-.125emX}}
\begin{document}

\title{Joint Multimodal Contrastive Learning for Robust Spoken Term Detection and Keyword Spotting}

\author{\IEEEauthorblockN{ Ramesh Gundluru}
\IEEEauthorblockA{\textit{Electrical Engineering} \\
\textit{Indian Institute of Technology}\\
Hyderabad, India \\
ee22m24p000001@iith.ac.in}
\and
\IEEEauthorblockN{Shubham Gupta}
\IEEEauthorblockA{\textit{Artificial Intelligence} \\
\textit{Indian Institute of Technology}\\
Hyderabad, India \\
ai22mtech12009@iith.ac.in}
\and
\IEEEauthorblockN{Sri Rama Murty K}
\IEEEauthorblockA{\textit{Electrical Engineering} \\
\textit{Indian Institute of Technology}\\
Hyderabad, India \\
ksrm@ee.iith.ac.in}


}
\maketitle

\begin{abstract}
Acoustic Word Embeddings (AWEs) improve the efficiency of speech retrieval tasks such as Spoken Term Detection (STD) and Keyword Spotting (KWS). However, existing approaches suffer from limitations, including unimodal supervision, disjoint optimization of audio-audio and audio-text alignment, and the need for task-specific models. To address these shortcomings, we propose a joint multimodal contrastive learning framework that unifies both acoustic and cross-modal supervision in a shared embedding space. Our approach simultaneously optimizes: (i) audio-text contrastive learning, inspired by the CLAP loss, to align audio and text representations and (ii) audio-audio contrastive learning, via Deep Word Discrimination (DWD) loss, to enhance intra-class compactness and inter-class separation. The proposed method outperforms existing AWE baselines on word discrimination task while flexibly supporting both STD and KWS. To our knowledge, this is the first comprehensive approach of its kind.
\end{abstract}



\section{Introduction}
Large-scale audio archives (e.g., podcasts, multimedia) continue to grow, yet they often lack transcriptions or rich metadata. Efficient search in such collections therefore requires spoken content retrieval techniques that index audio directly rather than relying on text\cite{7114229}. In this context, Spoken Term Detection (STD) and Keyword Spotting (KWS) are key tasks: KWS typically uses a written keyword (text query) to find matching utterances, whereas Query-by-Example STD (QbE-STD) matches a spoken query directly against speech content\cite{7114229} \cite{ram18_interspeech}. Conventional systems often cascade Automatic Speech Recognition (ASR) with text search\cite{7114229}, but ASR errors or resource constraints can severely limit performance. As a result, direct acoustic matching methods have gained attention as a way to bypass ASR entirely.

A common approach to QbE is Dynamic Time Warping (DTW)\cite{1163055} over frame sequences. In DTW-based methods \cite{settle2017querybyexamplesearchdiscriminativeneural}, both the query and search signals are aligned frame-by-frame to compute similarity. However, DTW is brittle in practice\cite{article}. It assumes clean, tightly cropped segments with matching endpoints\cite{dvornik2021dropdtwaligningcommonsignal}, making it sensitive to noise, and silence, while also suffering from computational complexity that scales quadratically with sequence length. To improve the suitability of DTW, several modifications have been introduced, such as Segmental DTW\cite{9413827}, Subsequence DTW\cite{Alshehri2019SubSequenceDTW}, and Non-Segmental DTW. Nevertheless, these approaches continue to face several challenges, including the lack of trainable parameters, high computational demands, sensitivity to background noise and speaker variability, and difficulty in determining appropriate matching thresholds.

To overcome DTW’s limitations, recent work has moved toward Acoustic Word Embeddings (AWEs): fixed-dimensional vector representations of variable-length audio segments. Under this paradigm, each word or phrase is mapped by a neural encoder to a point in a high-dimensional space where similarity can be measured by a simple distance metric (e.g., cosine). Early template-based embeddings (e.g., Levin et al.) encoded a word by its DTW distances to a set of reference templates\cite{settle2017querybyexamplesearchdiscriminativeneural}. More powerful neural AWEs have since been proposed. For example, Settle et al. train a Siamese LSTM\cite{settle2016discriminativeacousticwordembeddings} with a triplet loss so that instances of the same word project nearby in the embedding space. These discriminative AWE models have shown far better word discrimination than DTW methods. Other supervised approaches include multi-view RNNs (He et al.)\cite{he2016multi} that fuse different feature views into a single embedding space. In parallel, unsupervised seq2seq autoencoders have been explored: Chung et al.’s Audio Word2Vec\cite{chung2016audioword2vecunsupervisedlearning} uses an RNN encoder-decoder to reconstruct each segment, yielding embeddings that describes the sequential phonetic structures of words. Remarkably, this unsupervised model significantly outperformed DTW on a QbE-STD task at much lower computational cost. Kamper et al.\cite{peng2020correspondencevariationalautoencoderunsupervised} further extend this idea with a correspondence autoencoder (CAE): instead of reconstructing the same input\cite{chen2015query}\cite{cho2014learningphraserepresentationsusing}, the model is trained to reconstruct a different instance of the same word. By leveraging an unsupervised term discovery (UTD) system\cite{chung2016audioword2vecunsupervisedlearning} to mine same-word pairs\cite{jansen2013weak}, the CAE learns discriminative embeddings without labeled data. Finally, multilingual transfer approaches\cite{kamper2021improvedacousticwordembeddings} have shown that embedding models trained on high-resource languages can transfer knowledge to unseen low-resource languages.

Despite this progress, existing AWE methods still have gaps. Most models are trained unimodally \cite{ram18_interspeech,settle2017querybyexamplesearchdiscriminativeneural,peng2020correspondencevariationalautoencoderunsupervised,Jacobs2021AcousticWE} (using only audio-audio pairs or only audio-text pairs) and often treat audio–audio and audio-text objectives separately. As a result, they may excel at one task (e.g., QbE) but do not naturally support the other (e.g., text-STD). Most of the systems fail to exploit cross-modal input, limiting their flexibility. For broad spoken content retrieval, we desire a single model that can leverage both acoustic and lexical cues within a unified embedding space and can be trained even with minimal resources.

In this work, we propose a unified multimodal contrastive learning framework that addresses the limitations of existing AWE approaches. Our model learns a shared embedding space for both speech segments and text keywords by jointly optimizing: (1) a CLAP-style audio–text contrastive alignment, and (2) a DWD-inspired audio–audio discrimination objective. Specifically, speech segments and their text keywords are encoded using audio and text encoders. A contrastive loss brings matching audio–text pairs closer (as in CLAP)\cite{elizalde2022claplearningaudioconcepts}\cite{radford2021learningtransferablevisualmodels}, while a DWD loss enforces separation between embeddings of different spoken words \cite{Chantangphol2023EnhancingWD}. This multi modal training regime enhances robustness to speaker variation and background noise, enabling the model to focus on word identity across modalities. We evaluate our framework on the LibriSpeech\cite{panayotov2015librispeech} corpus using Average Precision (AP) as the primary metric, and demonstrate that our joint model consistently outperforms unimodal baselines across both STD and KWS tasks. Our main contributions are as follows:
\begin{enumerate}
    \item We introduce a novel joint multimodal contrastive learning framework that unifies speech and text in a shared embedding space, combining CLAP-style audio–text alignment with DWD-style audio–audio discrimination.
    \item It supports both STD and KWS, offering flexible cross-modal capabilities within a single model.
    \item We introduce a comprehensive and reproducible evaluation framework for AWSs, addressing inconsistencies in prior work by standardizing trial generation, explicitly handling IV/OOV splits, and releasing the evaluation recipe alongside our codebase.
\end{enumerate}

\begin{figure*}
    \centering
    \includegraphics[width=1\linewidth]{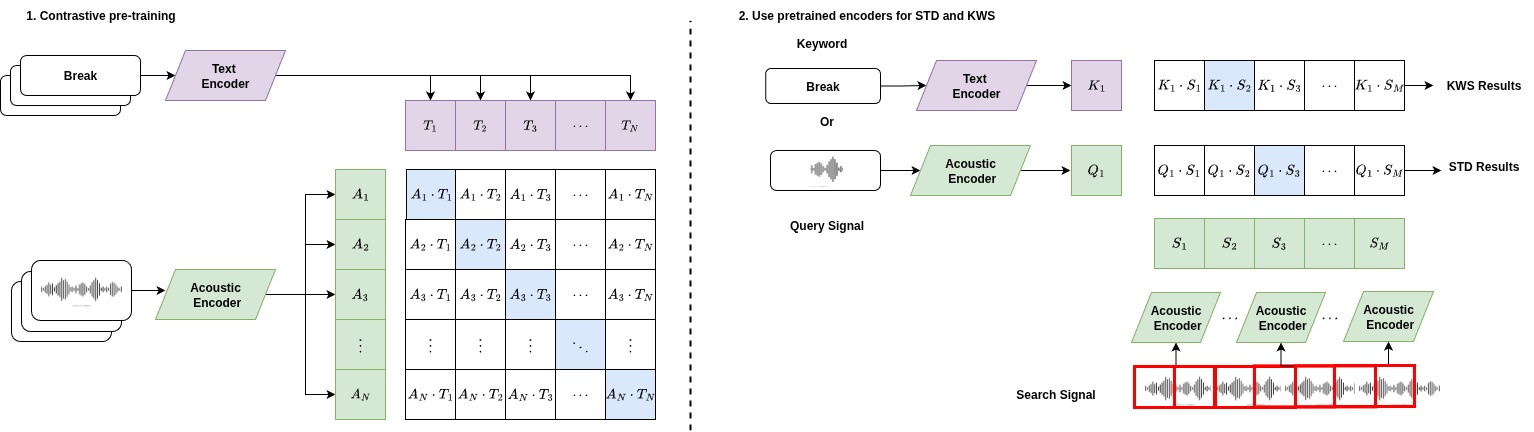}
    \caption{CLAP jointly trains audio and text encoders using contrastive learning to capture the similarity or dissimilarity between audio-text keyword pairs within a batch. During inference, the pretrained encoders generate embeddings for the spoken query and the text keyword. KWS or STD is then performed by computing the cosine similarity between these embeddings.}
    \label{fig:proposed_model}
\end{figure*}

\section{Related Works and Baselines}
\label{sec:baselines}



AWE methods have evolved across various modeling paradigms, ranging from fully supervised to unsupervised approaches. In this work, we benchmark our method against five representative systems that capture the diversity of these trends. These include Siamese networks with contrastive objectives, contrastive models trained with self-supervision, correspondence autoencoders with reconstruction objective, multi-view learning frameworks integrating audio and text, and architectures trained with word discrimination objectives. Together, these baselines reflect key developments in learning robust and transferable speech representations for STD and KWS.

\subsection{Siamese RNN}
Siamese RNN\cite{settle2017querybyexamplesearchdiscriminativeneural} is a BiLSTM-based model that learns fixed-dimensional representations from variable-length speech segments. It is trained on triplets of word segments—an anchor, a positive of the same word type, and a negative from a different word. The model minimizes a cosine hinge loss\cite{inbook}\cite{inproceedings} that encourages embeddings of the same word to be closer than those of different words:
\begin{equation}
\mathcal{L_{\text{hinge}}} = \max\left\{ 0, m + d_{\text{cos}}(f(a), f(p)) - d_{\text{cos}}(f(a), f(n)) \right\}
\end{equation}
where \( d_{\text{cos}} \) is cosine distance and $d_{\text{cos}}(u, v) = 1 - \frac{u^\top v}{\|u\|\|v\|}$
 and \( m \) is a margin. 

\subsection{Correspondence Autoencoder RNN}
Correspondence Autoencoder (CAE-RNN) is an unsupervised encoder-decoder RNN model designed to learn acoustic word embeddings using weak top-down constraints. Instead of reconstructing the same input as in standard autoencoders, CAE-RNN\cite{chung2016audioword2vecunsupervisedlearning}\cite{kamper2019truly} is trained on pairs of speech segments \( (X^{(a)}, X^{(b)}) \) where both segments are different instances of the same soken word. The encoder maps \( X^{(a)} \) to a latent representation \( \mathbf{z} \), which the decoder uses to reconstruct \( X^{(b)} \). The model minimizes the reconstruction loss:
\begin{equation}
\mathcal{L}(X^{(a)}, X^{(b)}) = \sum_{t=1}^{T_b} \left\| x^{(b)}_t - f_t(X^{(a)}) \right\|^2
\end{equation}
where \( f_t(X^{(a)}) \) denotes the decoder output at time \( t \) conditioned on the embedding of \( X^{(a)} \). 

\subsection{Multi-view Recurrent AWE}
Multi-View Recurrent AWE\cite{he2016multi} model jointly learns fixed-dimensional embeddings of acoustic speech segments and their corresponding orthographic (character-level) representations using a multi-view\cite{hermann2014multilingualdistributedrepresentationsword} contrastive learning framework. Two separate bidirectional LSTM encoders map an acoustic sequence \( x \) and a character sequence \( c \) into embeddings \( f(x) \) and \( g(c) \), respectively. These are trained such that matched pairs have lower cosine distance than mismatched ones. A typical objective is:
\begin{equation}
\mathcal{L} = \max\left\{ 0, m + d_{\text{cos}}(f(x^+), g(c^+)) - d_{\text{cos}}(f(x^+), g(c^-)) \right\}
\end{equation}
where \( d_{\text{cos}} \) is cosine distance, \( (x^+, c^+) \) is a correct acoustic-character pair. The framework also supports several alternative contrastive objectives by incorporating different types of mismatched negative pairs, including cross-view mismatches like \( (x^+, c^-) \) and \( (x^-, c^+) \), as well as same-view mismatches such as \( (c^+, c^-) \) and \( (x^+, x^-) \), enabling both cross-view and intra-view training. 

\subsection{Contrastive RNN}

Contrastive RNN\cite{Jacobs2021AcousticWE} learns fixed dimensional representation by optimizing a contrastive loss over multiple negative examples per anchor. Given a positive speech segment pair \( (X_a, X_p) \) and a set of negatives \( \{X_{n_1}, \ldots, X_{n_K}\} \), a shared encoder maps each segment to an embedding (e.g., \( \mathbf{z}_a = f(X_a) \)). The model minimizes a temperature-scaled cross-entropy loss that increases the cosine similarity between anchor and positive embeddings relative to negatives:
\[
\mathcal{L} = -\log \frac{\exp(\mathrm{sim}(\mathbf{z}_a, \mathbf{z}_p)/\tau)}{\sum_{j \in \{p, n_1, \ldots, n_K\}} \exp(\mathrm{sim}(\mathbf{z}_a, \mathbf{z}_j)/\tau)}
\]
where \( \mathrm{sim}(\mathbf{u}, \mathbf{v}) = \frac{\mathbf{u} \cdot \mathbf{v}}{\|\mathbf{u}\| \|\mathbf{v}\|} \) and \( \tau \) is a temperature parameter. 

\subsection{AWE with Deep Word Discrimination}
DWD\cite{Chantangphol2023EnhancingWD} loss is a supervised objective designed to enhance the quality of AWEs by enforcing both intra-class compactness and inter-class separability.  A detailed formulation of the DWD loss is provided in Section \ref{sec:method} of the paper.

\section{Methodology}
\label{sec:method}
Proposed multimodal pretraining framework for AWE learning is inspired by the CLAP model, that combines both audio and text modalities. We jointly train an audio encoder and a text encoder to project spoken queries and text keywords into a common semantic embedding space. The model learns to align semantically matched pairs while distinguishing mismatches using symmetric contrastive loss. We also used DWD loss to enhance AWEs by enforcing intra-class compactness and inter-class separability. The proposed framework is explained clealy in the following subsections. 

\subsection{Contrastive Audio Language Pretraining}

We extract the mel filter bank energy coefficients of the spoken query denoted by  \( X_a \in \mathbb{R}^{T \times F} \), where \( T \) is the number of time frames.
 and \( F \) is the number of coefficients. Similarly, the corresponding text keyword is denoted by \( X_t \in \mathbb{R}^{P \times V} \), where P is the number of phonemes in a keyword and V is the embedding dimension. We make sure that each training batch contains \( N \) such unique audio-text queries, denoted as \( \{X_a^i, X_t^i\} \) for \( i \in [1, N] \). For simplicity, we omit the superscript \( i \) and denote the entire batch as \( \{X_a, X_t\} \).

The audio inputs \( X_a \) and text inputs \( X_t \) are passed through modality-specific encoders: an audio encoder \( f_a(\cdot) \) and a text encoder \( f_t(\cdot) \), results in fixed dimensional representation for each example:
\begin{equation}
    \hat{X}_a = f_a(X_a), \quad \hat{X}_t = f_t(X_t)
\end{equation}


where \( \hat{X}_a \in \mathbb{R}^{N \times V} \) and \( \hat{X}_t \in \mathbb{R}^{N \times U} \) are the resulting audio and text representations, respectively, with \( V \) and \( U \) being their respective output embedding dimensions.

To bring both these representations in a shared multimodal embedding space of dimension \( d \), we apply learnable linear projections:
\begin{equation}
E_a = g_a(\hat{X}_a), \quad E_t = g_t(\hat{X}_t)
\end{equation}

where \( E_a, E_t \in \mathbb{R}^{N \times D} \), and \( g_a, g_t \) are the linear projection layers. All the representations are normalized to the unit norm.
The cosine similarity between audio and text embeddings is calculated and given by 
\begin{equation}
    C = exp(\tau) \cdot (E_t \cdot E_a^\top)
\end{equation}

where the temperature parameter \( \tau \), which controls the scaling of logits in the softmax, is learned during training as a log-parameterized scalar, thereby eliminating the need to tune it as a hyperparameter, and \( C \in \mathbb{R}^{N \times N} \) contains similarity scores between all audio-text pairs in the batch. The diagonal elements correspond to positive pairs, while the off-diagonal elements represent negative pairs. We use a symmetric contrastive cross-entropy loss over the similarity matrix: 
\begin{align}
    \mathcal{L}_{\text{audio}} &= -\frac{1}{N} \sum_{i=1}^N \log \frac{\exp(C_{i,i})}{\sum_{j=1}^N \exp(C_{i,j})} \\
    \mathcal{L}_{\text{text}} &= -\frac{1}{N} \sum_{j=1}^N \log \frac{\exp(C_{j,j})}{\sum_{i=1}^N \exp(C_{i,j})}
\end{align}



The losses are computed in both directions: $\mathcal{L}_{\text{audio}}$ aligns each audio with the correct text using a row-wise softmax, while $\mathcal{L}_{\text{text}}$ aligns each text with the correct audio using a column-wise softmax. This symmetric objective ensures balanced cross-modal alignment rather than favoring a single modality.  
The total cross modal loss is given by:

\begin{equation}
    \mathcal{L}_{at} = \frac{1}{2} \left( \mathcal{L}_{\text{audio}} + \mathcal{L}_{\text{text}} \right)
\end{equation}

This symmetric loss encourages paired audio and text embeddings to be close in the joint space, while pushing apart mismatched pairs, effectively learning discriminative representations suitable for audio-text keyword matching.

\subsection{Contrastive Audio Audio Pretraining}

While the audio-text contrastive loss effectively aligns audio and text embeddings in a shared space, it does not explicitly enforce structure within the audio embedding space itself. This lack of intra-class compactness and inter-class separation can limit the discriminative power of AWEs. To address this, we incorporate the DWD loss, which operates solely in the audio embedding space. It encourages embeddings of the same word to cluster around a class-specific centroid while simultaneously pushing them away from embeddings of other words. This complementary objective enhances the representational quality of the AWEs and improves keyword discrimination.

Given \( N \) distinct spoken queries in batch, sample  \( M \) positve instances for each example. The centroid \( \mathbf{c}_j \) for the \( j \)-th word class (excluding the current embedding \( \mathbf{e}_{ji} \)) is computed as:
\begin{equation}
    \mathbf{c}_j = \frac{1}{M-1} \sum_{\substack{m=1 \\ m \ne i}}^{M} \mathbf{e}_{jm}
\end{equation}

For each embedding \( \mathbf{e}_{ji} \), cosine similarities \( S_{ji,k} = \cos(\mathbf{e}_{ji}, \mathbf{c}_k) \) are computed against all class centroids. The total loss combines a softmax-based contrastive loss \( \mathcal{L}_{\text{sm}} \) and a contrastive centroid loss \( \mathcal{L}_{\text{cc}} \):
\begin{equation}
    \mathcal{L}_{\text{sm}} = -S_{ji,j} + \log \sum_{k=1}^{N} \exp(S_{ji,k})
\end{equation}
\begin{equation}
    \mathcal{L}_{\text{cc}} = \sum_{j=1}^{N} \sum_{i=1}^{M} \left[ (1 - S_{ji,j}) + \max_{\substack{k=1 \\ k \ne j}}^{N} S_{ji,k} \right]
\end{equation}
\begin{equation}
    \mathcal{L}_{\text{aa}} = \mathcal{L}_{\text{sm}} + \mathcal{L}_{\text{cc}}
\end{equation}
Since there are  M positives, we compute a $\mathcal{L}_{at}$ loss for every audio-text pair $N \times N$ and averaged the losses over these M matrices. We compute single $\mathcal{L}_{aa}$ loss across the batch of $N \times M$ audio embeddings.

To jointly optimize both objectives, we define the total loss as a weighted combination of the averaged CLAP loss and the DWD loss:

\begin{equation}
\mathcal{L}_{\text{total}} = \alpha_1 \cdot \frac{1}{M} \sum_{m=1}^{M} \mathcal{L}_{\text{at}}^{(m)} + \alpha_2 \cdot \mathcal{L}_{\text{aa}}
\end{equation}

Here, \( \alpha_1 \) and \( \alpha_2 \) are weighting factors that balance the contributions of the $\mathcal{L}_{at}$ loss and the $\mathcal{L}_{aa}$ loss, respectively.  This weighted formulation ensures both cross-modal alignment and audio-audio discriminability are learned simultaneously. The effect of varying the relative weights $(\alpha_1,\alpha_2)$  of $\mathcal{L}_{at}$ loss and $\mathcal{L}_{aa}$ loss on IV and OOV Word discrimination task is studied in TABLE \ref{tab:alphas} in Section \ref{sec:results}. For all subsequent experiments, we set the hyperparameters to $\alpha_1 = 0.1$  and  $\alpha_2 = 1$.

\section{Experimental Setup}
\subsection{Dataset and Preprocessing}
We utilize the LibriSpeech corpus for all experiments, specifically employing  the \textit{train-clean-100} subset ($\sim$100 hours) for training and the \textit{test-clean} subset ($\sim$5.4 hours) and \textit{test-other} subset ($\sim$5 hours) for evaluation. To obtain precise word-level boundaries, we apply the Montreal Forced Aligner (MFA) \cite{mcauliffe2017montreal}, which provides time-aligned annotations for each word in the audio files. Post-alignment, we filter word instances based on duration, retaining only those spanning between 0.5-2.0 seconds, a range commonly adopted in the literature to ensure the exclusion of excessively short or long keywords that may not be informative for embedding learning. This filtering results in a curated set of 26.2k unique anchor words, totaling 154k instances.

\subsection{Feature Extraction}
Audio segments are resampled to 16Khz and transformed into mel spectrograms using a 25 ms window size, 10 ms hop length, and 128 Mel filter banks, applying a Hann window to each frame to minimize spectral leakage. This representation captures the time-frequency characteristics essential for modeling speech signals.

\subsection{Model Architecture}
Our proposed model is benchmarked against five established baseline systems, each representing a distinct approach to AWE, as detailed in Section II. To ensure consistency across all models, we adopt a uniform architecture:
\begin{itemize}
    \item Audio Encoder: A 3-layer Bidirectional LSTM with a hidden size of 256, followed by a fully connected layer projecting to a 512-dimensional embedding space. 

\item Decoder (where applicable): Mirrors the encoder structure with a 3-layer LSTM and a hidden size of 256.

\item Text Encoder (for multi-view models and our proposed framework): A 3-layer Bidirectional LSTM with a hidden size of 256, followed by a fully connected layer projecting to a 512-dimensional space.
\end{itemize}

\subsection{Training Configuration}
All models were implemented in PyTorch and trained with a batch size of 128. We used the AdamW optimizer with a learning rate of 1e-3 and  weight decay of 1e-4 to regularize the training. Gradient clipping with a maximum norm of 1.0 was applied to prevent exploding gradients. The learning rate was scheduled using OneCycleLR with cosine annealing and a warmup phase covering 20\% of the total training steps. Training was performed for 30 epochs in total.  These settings are kept consistent across all experiments. For contrastive sampling, positive and negative pairs for each anchor were randomly selected.

\subsection{Evaluation Protocol}
We assess the intrinsic quality of the learned embeddings using a word discrimination task. Each test word segment is embedded, and the cosine distance between every pair is computed. By varying a distance threshold, we generate a precision-recall curve, with the AP serving as the evaluation metric. Recognizing the lack of standardized evaluation protocols in existing literature where datasets, trials generation, trials count, and in-vocabulary (IV) versus out-of-vocabulary (OOV) distinctions are often unspecified. We establish a comprehensive evaluation framework: \footnote{We released the trial generation recipe alongside our codebase to support reproducibility. The code will be available at \url{https://github.com/SIPLab-IITH/JMCL}.}
\begin{itemize}
    \item Trial Generation: For each anchor word with N instances, we consider all $\binom{N}{2}$ combinations as positive pairs. Negative pairs are formed by pairing each instance of an anchor word with all instances of non-anchor words.
    \item IV and OOV Splits: From the test set, words that also appear in the training set are classified as IV, while words not present in the training set are considered as OOV.
    \item Trial Counts: This methodology yields approximately 34.3 million word pairs for IV evaluation and 0.3 million for OOV evaluation.
\end{itemize}

\section{Results}
\label{sec:results}

\subsection{Word Discrimination Performance}
Table \ref{tab:my_label} summarizes the performance of word discrimination for all baselines and our proposed model evaluated on IV and OOV keywords from the LibriSpeech \textit{test-clean} split. In the Acoustic view,  when embeddings are produced solely from speech segments, our model achieves 85.05 \% AP on IV words and 94.06 \% on OOV words. Notably, CLAP by itself (with no DWD loss) already attains 74.83 \% IV and 91.68 \% OOV, but the addition of the DWD loss consistently boosts acoustic performance by approximately 10\% on both IV and OOV sets. This confirms our hypothesis that combining audio–text alignment (CLAP) with audio–audio discrimination (DWD) yields more compact intra-class clusters and larger inter-class margins, thereby improving pure acoustic word discrimination. 

\begin{table}[h]
    \centering
    \caption{Word discrimination task performance (Average Precision \%) from Acoustic and Cross views for IV and OOV keywords.}
    \label{tab:my_label}
    \begin{tabular}{c|c|c|c}
        \hline
        \textbf{View} & \textbf{System} & \textbf{IV} & \textbf{OOV} \\
        \hline
        \multirow{5}{*}{\textbf{Acoustic}} 
        & CAE-RNN          & 7.24  & 43.64 \\
        & Siamese-RNN      & 57.12 & 82.84 \\
        & Contrastive-RNN  & 74.94 & 77.22 \\
        & A2E-DWD          & 72.15 & 79.83 \\
        & Multiview-RNN    & 55.47 & 81.72 \\\cline{2-4}
        & CLAP             & 74.83 & 91.68 \\
        & CLAP + DWD       & \textbf{85.05}    & \textbf{94.06} \\
        
        \hline
        \multirow{3}{*}{\textbf{Cross}} 
        & Multiview-RNN    & 98.66 & 99.46 \\\cline{2-4}
        & CLAP             & 98.66 & 99.46 \\
        & CLAP + DWD       & \textbf{98.66} & \textbf{99.46} \\
        \hline
    \end{tabular}

\end{table}

\begin{table*}[ht]
\centering
\small
\caption{Performance in terms of Equal Error Rates (EER \%) of different systems across vocabulary types and segment sizes for both STD and KWS on \textit{test-clean} and \textit{test-other} datasets.}
\vspace{2mm}
\label{tab:precision}
\resizebox{\textwidth}{!}{%
\begin{tabular}{c|l|cccc|cccc|cccc|cccc}
\hline
\multirow{3}{*}{\textbf{Task}} & \multirow{3}{*}{\textbf{System}} 
& \multicolumn{8}{c|}{\textbf{test-clean}} & \multicolumn{8}{c}{\textbf{test-other}} \\
\cline{3-18}
& & \multicolumn{4}{c|}{\textbf{IV}} & \multicolumn{4}{c|}{\textbf{OOV}} 
& \multicolumn{4}{c|}{\textbf{IV}} & \multicolumn{4}{c}{\textbf{OOV}} \\
\cline{3-18}
& & 0.2 & 0.3 & 0.4 & 0.6 & 0.2 & 0.3 & 0.4 & 0.6 
  & 0.2 & 0.3 & 0.4 & 0.6 & 0.2 & 0.3 & 0.4 & 0.6 \\
\hline

\multirow{7}{*}{STD} 
 & CAE-RNN             & 28.01 & 26.96 & 34.67 & 53.60 & 44.71 & 41.40 & 33.75 & 34.52 & 39.42 & 36.34 & 41.38 & 46.44 & 35.76 & 31.66 & 29.94 & 40.30 \\
 & Siamese-RNN         & 22.70 & 21.71 & 30.00 & 46.82 & 39.05 & 31.01 & 22.88 & 27.71 & 34.34 & 28.83 & 36.75 & 46.04 & 37.33 & 29.86 & 24.59 & 44.80 \\
 & Contrastive-RNN     & 20.64 & 20.24 & 29.11 & 45.62 & 38.70 & 32.56 & 34.57 & 30.10 & 30.35 & 27.21 & 25.86 & 34.00 & 34.76 & 35.59 & 37.69 & 43.46 \\
 & Multi-view-RNN      & 22.11 & 22.72 & 26.78 & 37.77 & 42.92 & 35.51 & 30.49 & 32.17 & 33.26 & 30.33  & 32.28 & 34.60 & 31.62 & 28.79 & 27.51 & 41.15 \\
 & A2E-DWD             & 37.67 & 25.45 & 37.81 & 48.81 & 35.21 & 37.62 & 32.34 & 31.42 & 40.62 & 29.64 & 35.27 & 42.61 & 47.49 & 42.25 & 39.65 & 48.05 \\ \cline{2-18}
 & CLAP                & 20.17 & \textbf{16.34} & 19.24 & 33.12 & 38.87 & 31.75 & \textbf{18.00} & 18.34 & 29.20 & 24.68 & \textbf{23.40} & 31.97 & 30.16 & 22.89 & \textbf{18.21} & 32.33 \\
 & CLAP + DWD          & 20.24 & \textbf{15.71} & 21.08 & 33.89 & 30.53 & 25.84 & \textbf{18.35} & 19.35 & 27.41 &\textbf{ 22.20} & 22.95 & 32.04 & 28.29 & \textbf{21.03} & 18.56 & 33.81 \\
\hline
\multirow{3}{*}{KWS} 
 & Multi-view-RNN      & 42.95 & 44.32 & 45.10 & 45.24 & 37.74 & 40.23 & 39.73 & 35.21 & 37.84 & 41.05 & 40.21 & 38.06 & 51.25 & 47.60 & 48.83 & 53.38 \\ \cline{2-18}
 & CLAP                & 39.36 & \textbf{36.21} & 37.21 & 40.36 & 39.98 & 38.10 & 40.23 & 41.06 & 41.01 & 41.65 & 43.84 & 43.77 & 61.40 & 55.37 & 53.52 & 54.48 \\
 & CLAP + DWD          & 49.01 & 51.40 & 48.13 & 43.36 & 40.02 & 39.00 & 36.85 & 36.00 & 43.76 & 46.30 & 45.58 & 43.45 & 57.57 & 53.45 & 55.45 & 56.03 \\
\hline
\end{tabular}}
\end{table*}

In the Cross-view where models embed both speech and text so that an audio embedding can be directly compared to a text embedding, CLAP achieves 98.66\% AP on IV and 99.46\% on OOV, matching the Multi-view-RNN baseline. Importantly, adding DWD to CLAP does not degrade performance, and AP remains at 98.66\% / 99.46\% for IV/OOV splits. In this cross-modal scenario, DWD provides negligible additional benefit, as expected: DWD's audio–audio discrimination is redundant when the primary training and evaluation objective is audio–text alignment. Nonetheless, CLAP + DWD does not sacrifice any cross-view quality on word discimination task.

Across all methods, OOV performance exceeds IV performance. Two factors help explain this counterintuitive result. First, the IV set contains more total word-pair combinations, roughly 34 million IV pairs versus 0.3 million OOV pairs. So, the IV AP estimate is based on a far denser evaluation grid. Second, each IV anchor has, on average, 5.64 instances in the test split (and was explicitly seen during training with at least 100 instances in the train split), whereas OOV anchors have only about 1.4 instances on average. In effect, the smaller, sparser OOV evaluation set yields fewer hard negative and positive trials per anchor, inflating AP. In contrast, the IV set’s richer sampling produces a more challenging discrimination task.

\subsection{Query-by-Example STD and KWS}

Table \ref{tab:precision} reports Equal Error Rates (EER) for both STD and KWS under varying segment sizes for IV and OOV keywords on \textit{test-clean} and \textit{test-other} datasets. AWEs require segmenting the continuous test audio into fixed-length segments (with a fixed hop) for STD and KWS tasks. The boundaries of the window can cut through target words, omit parts of words, or capture multiple words in a single window. Consequently, all models exhibit Higher EER relative to their word-discrimination AP results.  In the STD task, our model consistently achieves the lowest EER across most window sizes on \textit{test-clean} dataset. For IV queries, it obtains the best overall performance with an EER of 15.71\% at a 0.3-second window. While CLAP reports the lowest OOV EER of 18\%, our model closely follows with 18.35, outperforming other baselines. 


On the more challenging \textit{test-other} set, all methods suffer higher EER due to increased noise and speaker variability. At a 0.4-second window, CLAP yields  23.40\% EER for IV and 18.21\% EER for OOV. CLAP+DWD improves EER to  22.20\% for IV,  but shows only marginal gains on OOV queries. Our model maintains stronger performance, with consistently lower EERs across segment sizes, and narrower gaps between \textit{test-clean} and \textit{test-other} for STD tasks. These results confirm that our joint objective is more robust to mismatched conditions, yielding embeddings that remain discriminative even under noisy and diverse speech scenarios.

In the KWS task on \textit{test-clean}, CLAP performs better than the baseline. However, the inclusion of the DWD loss in CLAP+DWD slightly degrades performance, yielding EER of 43.36\% and no improvement on OOV queries. This drop is attributed to the DWD loss, which operates solely on audio embeddings and does not enhance cross-modal alignment.

\subsection{Qualitative Analysis}

\begin{figure}
    \centering
    \includegraphics[width=0.6\linewidth]{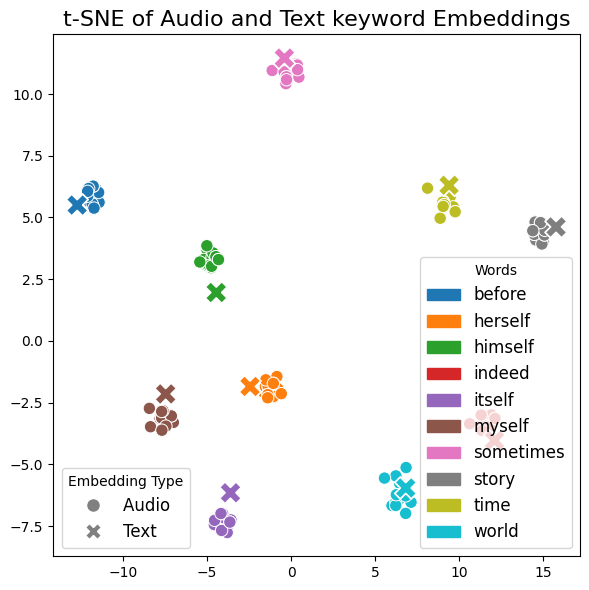}
    \caption{t-SNE visualization of audio and text keyword embeddings. Audio embeddings (circles) are obtained from 10 spoken instances of 10 selected keywords, while text embeddings (crosses) represent the corresponding phonetic queries. Each color corresponds to a different keyword.}
    \label{fig:tsne}
\end{figure}

We utilized t-SNE to project a subset of audio and text keyword embeddings into a two-dimensional space, yielding a visual representation of the joint embedding space, as shown in Figure \ref{fig:tsne}. The resulting plot demonstrate that audio embeddings for each query closely cluster around their corresponding text embeddings, validating the effectiveness of the CLAP module in aligning cross-modal features. Additionally, distinct anchor clusters exhibit clear separation, indicating that the DWD loss effectively enforces robust inter-class discrimination within the audio embedding space. 

\begin{figure}[h]
    \centering
    \includegraphics[width=1\linewidth]{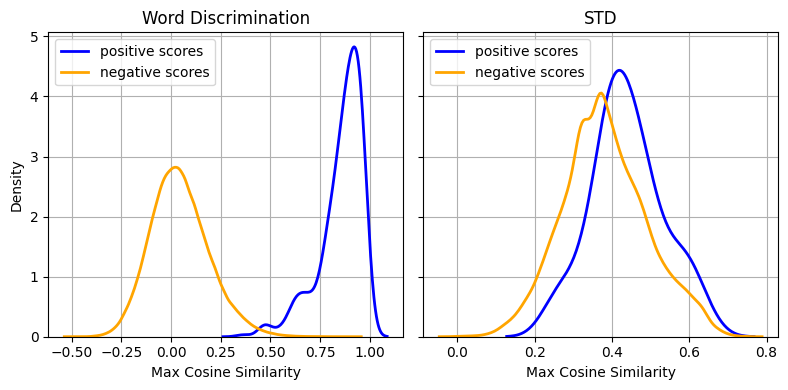}
    \caption{Kernel Density Estimation (KDE) plots of maximum cosine similarity
scores for positive  and negative utterances 
using fixed window and ground truth locations.}
    \label{fig:cosine_scores}
\end{figure}

We analyzed Kernel Density Estimates (KDE) of cosine similarity scores for positive and negative pairs in two conditions: (1) a word discrimination setup using accurately aligned segments obtained via MFA, and (2) STD setup using fixed-length windowed segments (0.3 sec length, 0.15 sec hop).  In the MFA-aligned case, positive pairs show high similarity scores (mean $\approx$ 0.89) with clear separation from negative pairs (mean $\approx$ 0.07), resulting in a well-defined margin as illustrated in Fig \ref{fig:cosine_scores}. However, with windowed segmentation, the positive pair similarity distribution shifts lower (mean $\approx$ 0.45) and becomes more dispersed due to windows often cutting through keywords or capturing irrelevant content. This misalignment leads to degraded discriminability and higher EER in KWS and STD tasks

\begin{table}[]
    \centering
    \caption{Effect of Varying Loss Weights $(\alpha_1,\alpha_2)$  for $\mathcal{L}_{at}$  and $\mathcal{L}_{aa}$  on IV and OOV Word Discrimination Performance}
    \begin{tabular}{c|c|c}
        \hline
        \textbf{Loss Weights} &  \textbf{IV} & \textbf{OOV}  \\
        \hline
         $\alpha_1$ = 1.0, $\alpha_2$ = 0.1  & 78.40 & 72.88 \\
         $\alpha_1$ = 1.0, $\alpha_2$ = 0.5  & 80.45 & 68.86 \\
         $\alpha_1$ = 1.0, $\alpha_2$ = 1.0  & 82.37 & 87.21 \\
         $\alpha_1$ = 0.1, $\alpha_2$ = 1.0  & \textbf{85.05} & \textbf{94.06} \\
         $\alpha_1$ = 0.5, $\alpha_2$ = 0.1  & 84.48 & 91.41 \\
         \hline
    \end{tabular}

    \label{tab:alphas}
\end{table}

\section{Conclusion}
In this work, we introduced a unified multimodal contrastive learning framework that simultaneously aligns audio and text while enforcing audio–audio discriminability, yielding a single embedding space for both STD and KWS. Our model consistently outperformed unimodal baselines on LibriSpeech, especially in noisy and speaker-variant conditions, and exhibited strong generalization to unseen (OOV) words. Qualitative analyses (t-SNE and KDE plots) confirmed that the embeddings form tightly clustered, well-separated word classes across modalities. We also release a standardized, reproducible evaluation framework addressing prior inconsistencies.



\bibliographystyle{IEEEtran}
\bibliography{IEEEabrv,references}

\end{document}